# PATTERNS OF ICT USAGE IN DISASTER IN SAMOA

Ioana Chan Mow, National University of Samoa, i.chanmow@nus.edu.ws

Agnes Wong Soon, National University of Samoa, a.wongsoon@nus.edu.ws

Elisapeta Maua'i, National University of Samoa, e.mauai@nus.edu.ws

Ainsley Anesone, National University of Samoa, a.anesone@nus.edu.ws

**Abstract:** The study discussed in this paper focuses on ICT use during disasters in Samoa and is a replicate of a study carried out in 2015. The study used a survey to explore how Samoan citizens use technology, act on different types of information, and how the information source or media affects decisions to act during a disaster. Findings revealed that traditional broadcasting were still the most prominent, most important, and still predominate in early warning and disaster response. However, there were now increasing usage of mobile and social media in disaster communications. Findings also revealed that people trust official reporters the most as source of information in times of crisis. The intent is that findings from this study can contribute to a people-centred approach to early warning and disaster providing empowerment to affected individuals to act in a timely and appropriate manner to ensure survival in times of disaster.

## 1. INTRODUCTION

While technology access has expanded worldwide, there is still limited research about how people use the technology available to them during crisis, and what the effects of varying socio-political and economic contexts are on the level of trust people place on information from different sources and technical mediums. The study discussed in this paper focuses on ICT use during disasters in Samoa at the individual level and is a replicate of the original study carried out in 2015 (Mow, Shields, Sasa, & Fitu, 2017).

The objectives of the research were to resolve the following questions:

**How can Samoans use their mobile and internet devices during times of crises and natural disasters?**

**How do Samoan citizens act on different types of information and how does the source of information affect their decisions to act?**

Attempts were also made to compare the findings to the 2015 study. The aim of the second study was to determine if any significant changes had occurred in the last 5 years as well as the added interest of early warning systems in the light of the current pandemic. In Samoa disasters such as cyclones and flooding are common occurrences particularly during the wet season. Timely warning of approaching cyclones and the imminence of floods saves lives and the use of technology coupled with timely action based on trusted sources of information are critical factors and hence the motivation for this research. The intent is that findings from this study can contribute to a people-centred approach to early warning and disaster providing empowerment to affected individuals to act in a timely and appropriate manner to ensure survival in times of disaster. Early warning is defined as 'the provision of timely and effective information, through identified institutions, that allows individuals exposed to a hazard to take action to avoid or reduce their risk and prepare for effective response' (UNISDR, 2004). A people-centred early warning system comprises of four interacting elements (i) risk knowledge, (ii) monitoring and warning service, (iii) dissemination and communication and (iv) response capability (UNISDR, 2004). As pointed out by Twigg (2002), the human factor in early warning systems is highly significant and failures attributed mainly to the communication and preparedness elements (Basher, 2006). At the root of all this is the need for the provision to communities of information which can empower them to make the right decisions to protect themselves and respond accordingly. Factors such as risk awareness, personal trust to the government and mass media, involvement in the community and social networks can influence their preferred warning information source (Lindell & Perry, 2012; Rahayu, Comfort, Haigh, Amaratunga & Khoirunnisa, 2020).

The main area of interest in this research is to understand how citizens can use their mobile and internet devices during crises and natural disasters. Since communication infrastructure is a "magnifier of human intent," there is a relationship between social capital, good governance, and how people act on different streams of information. Such a study is also timely in the light of the current pandemic with its heavy reliance on technology and also the need to investigate how information can be transmitted vertically and laterally in times of such a crisis.

## 2. LITERATURE REVIEW

While technology access has expanded worldwide, there is still limited research about how people use the technology available to them during crisis, as well as what the effects of varying socio-political and economic contexts are on the level of trust people place on information from different sources and technical mediums. While the commercial and economic benefits of increased mobile phone and data connectivity have been evident to agencies such as the World Bank and the telecommunications industry since the early 2000s, the benefits of these technologies with regards to governance and crisis response have only started to emerge in the last eight to ten years (Eggleston, Jensen & Zeckhauser 2002; Kirkman, Cornelius, Sachs, & Schwab 2002).

In discussing ICTs and development it is important to differentiate between social and technological research. A great deal of research and development has gone into software and hardware for development and crisis response. Instances include the Qatar Computer Research Institute (QCRI) which is leading efforts to automate the tracking and coding of Tweets from disaster zones (Meier, 2013) and Ushahidi who has developed a brick-sized device that acts as a remote internet hotspot (Rotich, 2017). Furthermore, data science continues to expand our understanding of how to use the vast amounts of data generated via ICTs (Crawford, 2013).

The current study focuses on the human side of technology use during crises, using surveys to ascertain what information technologies people choose to use when making a decision to act during a crisis.

Letouze, Meier and Vinck (2013) argued that big data (and by extension the technologies that allow people to contribute to big datasets) can be packaged and used in such a way that it empowers communities to develop their own local-level responses to crisis, since larger government and governance organizations often cannot react quickly to emerging risks. This concept is mirrored in an earlier paper by Meier (2008), critiquing the problems with traditional conflict early-warning systems; the methods that are used for early warning are too aggregate to provide information to policy makers and first responders quickly enough to prevent violence.

Crises present a variety of collective action and decision-making problems, involving high risk and low information. In crises two factors are key in collective action problems: one is group size and the other is group cohesion (Olson, 1965). ICTs help alleviate both of these problems, because they can broadcast to a very large audience with little difference in cost (Lupia & Sin, 2003). The transmission costs of an individual text message versus a text message sent to hundreds of recipients is negligible. In a social networking platform like Twitter, thousands of people can receive and rebroadcast a message within minutes at no cost to the original sender (except for the overhead of having an internet connection). ICTs, because of their technical attributes, take care of the problems related to population size, cost, and regularity of information transmission. Information is a critical resource in times of disaster allowing responders to effectively manage a disaster and those affected to best adapt to the threat (Steelman, McCaffrey, Velez & Briefel, 2015). It is in these situations where ICTs become useful, since they can be used to share large amounts of data laterally and vertically very quickly.

This leaves the socio-political question of what information mediums and sources people trust and will act on. As Gurstein (2014) pointed out, "..what is particularly important here is the significance not simply of the availability of information but also of the capacity to identify what information is important, who the information would be important to, and how to bring that information to the attention of those for whom it will be important and useful".

But ICTs do not cause people to act; people take action when they trust the information they receive, and trust is a socio-political outcome not a technology-driven one. Trust is a key element which influences the response of a receiver of information in times of disaster (Renn & Levine, 1991; Kasperson & Stallen, 1991).

Until recently scant attention has been paid to what sources of information recipients turn to, find useful or trust in times of disaster (Steelman et al., 2015). An understanding of what sources and media of information people trust and regard as useful and important will result in more effective communication in early warning and disaster. These are the aspects of disaster response which are evaluated in this research. The research investigates what sources and medium of information responders or people trust in times of disaster; what sources and media they will act upon and consider important.

In Samoa, technology is used in disaster and relief management in various ways. Alerts forwarded from the Tsunami centre in Hawaii and early warning from the Emergency Managers Weather Information Network (EMWIN) are received by Disaster management organisation (DMO), Meteorology section via mobile, email and by fax. These alerts are then disseminated to selected citizens in every village via a range of modern and traditional media such as pre-programmed SMS, radio, tv and email. These will then give rise to warning signals such as use of church bells, school bells and also continuous sound of sirens to warn the public of imminent and approaching disaster.

## 3.    METHODOLOGY

The proposed study is quantitative in nature. A survey was used to collect data. The survey instrument in the form a questionnaire consisted of the following sections:

  i. Respondent details
 ii. Ownership of phones and accessibility to selected social sites
iii. News and information
 iv. Emergency information-(e.g.,) cyclone, earthquake, tsunami
  v. Taking action during an emergency (e.g.,) cyclone, earthquake, tsunami
 vi. Importance of information sources

(Please refer to the English and Samoan version of the questionnaire attached.)

### 3.1   Sample

Sample selection was based on both population size and geographical location to ensure representativeness. The proposed study used the geographical subregions based on the Samoa Bureau of Statistics (SBS) census divisions, where Samoa is divided into 4 subregions: i) Savaii, ii) Urban Upolu, iii) Northwest Upolu and iv) the Rest of Upolu. The survey was conducted on a sample of 400 households based on 1 participant interviewed per household. To ensure a representative sample indicative of population density and geography, proportionate sampling was used at the level of subregions based on SBS 2016 Census data. Hence the sample of 400 comprised of 89 from Savaii, 76 from Urban Upolu, 142 from North West Upolu and 93 from the Rest of Upolu. Within each subregion, convenience sampling was used to obtain the necessary samples within districts and villages.

|  | Total population SBS 2016 Census | % | Weighted Sample by subregion |
|---|---|---|---|
| Samoa | 195,979 |  |  |
| Apia Urban Area | 37,391 | 19.1 | 76 |
| North West Upolu | 69,376 | 35.4 | 142 |
| Rest of Upolu | 45,652 | 23.3 | 93 |
| Savaii | 43,560 | 22.2 | 89 |

Table 1. Weighted sample of 400 based on 4 subregions of Samoa

### 3.2 Procedures

The survey team administered the survey to respondents in the 4 subregions. Eligible respondents for the survey were members of the surveyed household in the age range 18 to 40. The survey and the consent form were made available in both English and Samoan. For each subregion the survey team used convenience sampling and surveyed people in key convenience locations such as shops and community centres. After signing the consent form, each participant was questioned by a member of the survey team and responses recorded on the survey form.

### 3.3 Data Analysis

Data included information on trust and use of information from different sources and technical mediums (e.g. mobile phone, social media, etc.).

Statistical analysis was conducted using SPSS with the data primarily being descriptive statistics.

## 4. RESULTS AND DISCUSSION

The results of the analysis are reported in the sections in the order within which they are located in the questionnaire. In the case of multiple response items, the findings are reported as percentages or proportion of the total sample (400).

### 4.1. Nature of Study Participants

There were 400 respondents for this survey from the 4 subregions of Samoa with proportions reflecting 2016 Census proportions by subregion. The sample was selected across 38 districts and 141 villages. Respondents were in the age range 18 to 75 with an average age of about 31. Sample were predominantly female and consisted of 36.2% male and 63.8% female. In terms of marital status 54% were single, 41.5% married, 3% divorced /separated and 1.5% widowed. In terms of employment status, sample were predominantly employed (42.5%), 32% students, 10% domestic duties, 10% self-employed, 0.5 % special needs.

### 4.2. Ownership of phones and accessibility to selected social sites

Results of the survey indicated that 94% of the respondents owned a mobile phone. This figure is about the same as the World Bank national statistic (2020) on mobile ownership of 90%. Of these mobile phone owners, 82.5% owned smart phones. The findings indicate that mobile phone ownership remain relatively high and similar to the 2015 figure of 91% but a massive increase in ownership of smartphones from 34% in the 2015 survey to 82.5% in the 2020 survey. This implies that any emergency services can now target smartphones. A possible cause of this is the affordability and accessibility of mobile phones. With the high rate of mobile ownership and smartphones, this also implies that use of SMS and cell broadcasting and other mobile apps can be viable and cost-effective forms of dissemination of emergency warnings and notices.

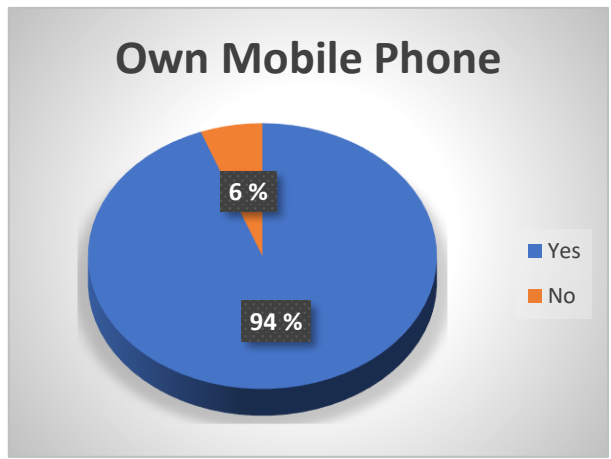

**Figure 1. Ownership of mobile phones**

In terms of mobile providers, results indicated that Digicel was the dominant provider at 52% followed by Vodafone 32% and with 15 % (23%) indicating service from both providers. Compared to the 2015 survey, Digicel was still the dominant provider but Vodafone (previously Bluesky) had now doubled its market share compared to 15% in 2015. This finding has valuable implications on decisions on SMS and cell broadcasting services.

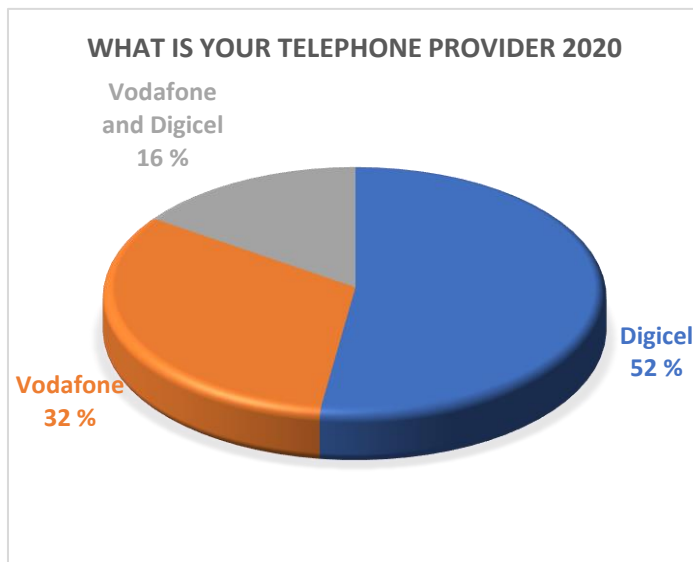

**Figure 2. Telephone providers**

About 89% of the respondents with cell phones indicated that they use their cell phone for Internet access compared to 56% in the 2015 study. An evaluation of how people access Internet from home indicated that 89% access through mobile phone, 24% use home computers and only 8% use tablets. Again, the findings indicate a sizable increase in access by mobile phone from 52% in 2015 to 89% in 2020.

An evaluation of the use of applications revealed that about 88% use IPchat applications, 90% use Facebook and only 11% use Twitter. These figures indicate a huge increase (almost triple) in the use of IPchat applications from 26% in 2015 to 88%, as well as 65% in 2015 to 90% in 2020 for Facebook usage. The large proportion of IPchat application users and Facebook users indicate that these types of social media may be used for dissemination of disaster warnings and response information.

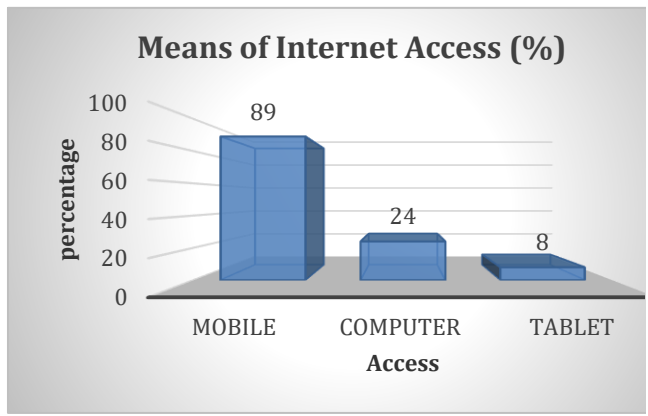

**Figure 3. Means of Internet Access**

### 4.2.1. Who People Trust in an Emergency

In trying to understand collective action problems, one focus of this study was to determine who people trust during an emergency. Results indicated that 95% trusted government and 84% trusted that the Matai or council of chiefs will provide them with the necessary emergency relief. These figures indicate an increase in trust in both government and matai from the 2015 survey but also that people continue to place a high level of trust in government to provide them with the necessary emergency relief during disasters, as well as the village leaders (Matai).

### 4.2.2. Who do People Turn to for help

When asked as to whom they turn to for help, respondents gave a range of responses. The most common responses were family (31%) followed by government (13%) and then the Disaster Management Office (DMO) (10%). When asked for reasons given for their choice of who to turn to, majority indicated "trust" (18%)," those that act fast" (14%) and that their choice was due to that" it was their job to act" (13%). Such responses are useful in guiding decisions on key agents for disaster and emergency relief response. It is also interesting to note that participant responses to this item, were shaped and conceptualized in the light of recent emergency events such as the measles epidemic and the current COVID pandemic.

### 4.3. News and Information

This section of the survey aimed to determine sources of news and information used by people. When asked from whom they get their news from, most respondents indicated professional reporters (83%), friends (50%), family matai (46%), government (39%) and mayor (Pulenuu) (20%), as the main source of news and information. Professional reporters refer to official media personnel such as radio, television and newspaper reporters. There was an overall increase compared to 2015 across all categories with professional reporters still the most popular source of news.

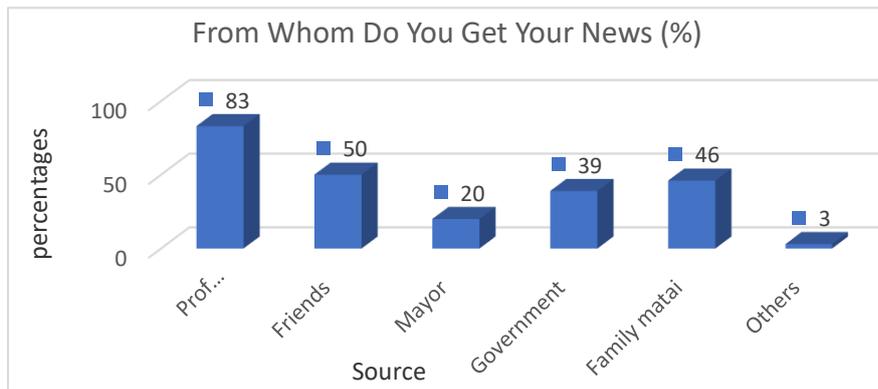

**Figure 4. Source of news**

### 4.3.1. Sources People get News from

In terms of what sources people get news from, findings indicated a predominance of tv with 84% indicating TV1 as their news source, 44% TV3, 20% EFKS TV. In terms of radio 62% source news from Talofa FM, 51% from Radio 2AP, 21% from Magik FM. There was also an increase in Mobile SMS as a news source to 57%, Internet at 40% and newspaper at 21%. These findings indicate a shift of predominance to both tv and radio as well as increased prominence of the use of mobile SMS and the Internet as sources for news. These findings are similar to studies by Becker (2004), Taylor et al., (2007) and Cretikos et al,. (2008) which indicated the predominance of mass media such as radio and television as a source of information.

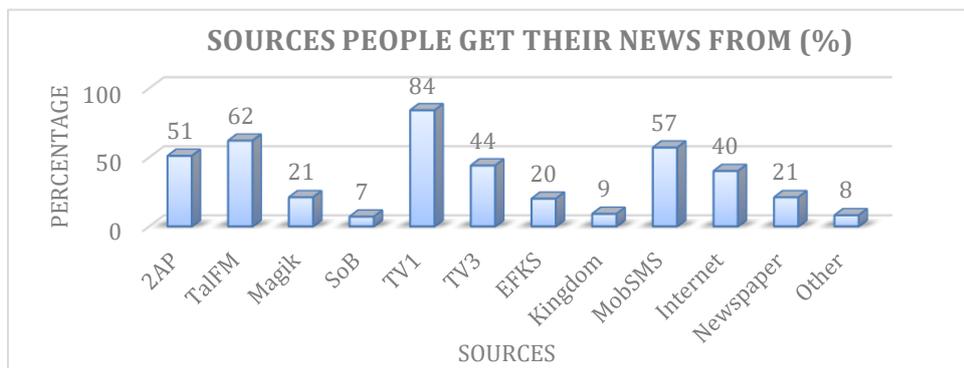

**Figure 5. Sources People Get News From**

### 4.4. Emergency Information

### 4.4.1. Whose information People trust in an emergency

Respondents were asked about whose information they trust during an emergency. Responses indicated 45% trusted information from professional reporters, 33% trust information from government, 28% trust information from the family matai[1] and 13% trust friends as the main source of information they trust in an emergency. Only 10% of the respondents indicated the village Pulenuu as a trusted source of information. These findings suggest a change from 2015 where professional reporters and family matai were the main sources of trusted information. It also suggests an increase in trust in government for providing information during an emergency.

---

[1] Family chief or head of the family

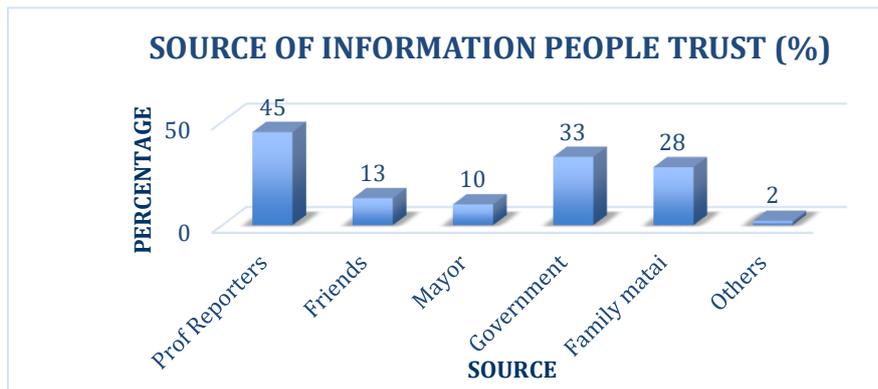

Figure 6. Whose information do you trust

### 4.4.2. What sources of media do people trust in an emergency

In terms of what sources of media were trusted in times of emergency, there were a range of responses. Findings indicate the 5 main sources people trust are TV1 (49%), Talofa FM radio (35%), Radio 2AP (34%) Internet (30%) and Mobile SMS (29%). These findings seem to suggest a shift towards both radio and tv and also increasingly now the Internet and Mobile SMS in terms of sources people trust in an emergency.

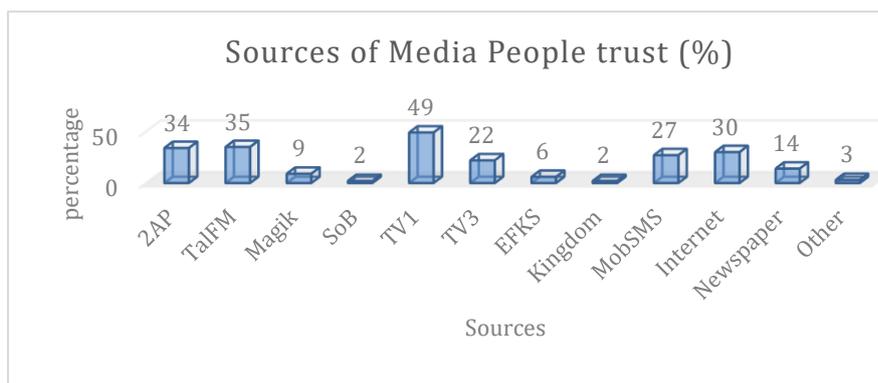

Figure 7. What sources of information do you trust

### 4.5. Taking Action during an Emergency

### 4.5.1. Whose information people act upon during an emergency

In emergency situations it is not sufficient for people to just receive information but also to act upon information given in a timely manner in order to save lives and prevent hazardous impacts of disaster. Results of the survey indicated that during an emergency people act primarily upon information from professional reporters (37%), family chief (matai) (34%) followed by government (33%), the village pulenuu[2] at 14% and friends (9%). These findings are similar to the 2015 findings.

---

[2] Village mayor

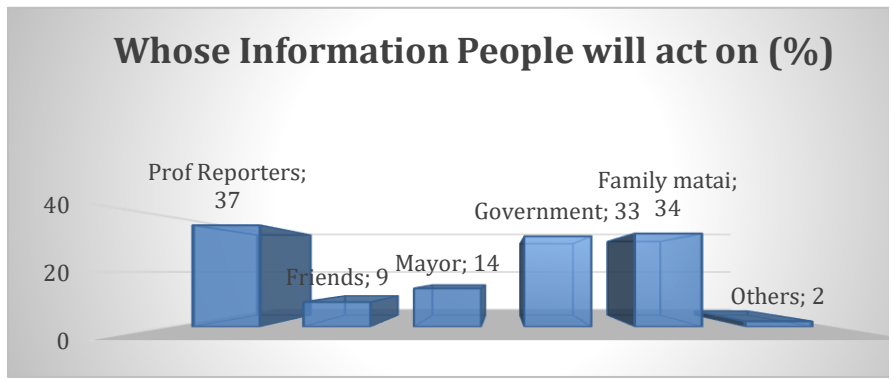

**Figure 8. Whose information do people act upon**

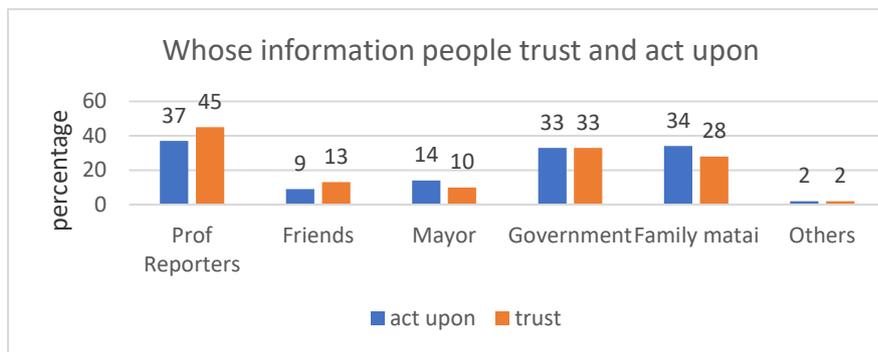

**Figure 9. Whose information people trust and act on**

Furthermore, a graph of whose information people trust and act upon (Figure 9), indicate that proportions of these two are very similar, thus indicating that people will act upon information from sources that they trust.

### 4.6. Importance of Information Sources

#### 4.6.1. What sources of information people act upon

Investigation of what sources of information people act upon during an emergency using percentages of total sample indicated that findings follow closely the proportions of what sources of information people trust during an emergency. People acted upon information from sources of media that they trust. Hence, as indicated earlier, findings indicate the 5 main sources people trust and act upon are TV1 (44%), Talofa FM radio (33%), Radio 2AP (32%) Internet (24%) and Mobile SMS (25%). These findings seem to suggest predominance of both radio and tv and also increasingly now the Internet and Mobile SMS in terms of sources people trust and act upon in an emergency. This suggests that people will act upon information from sources of media that they trust.

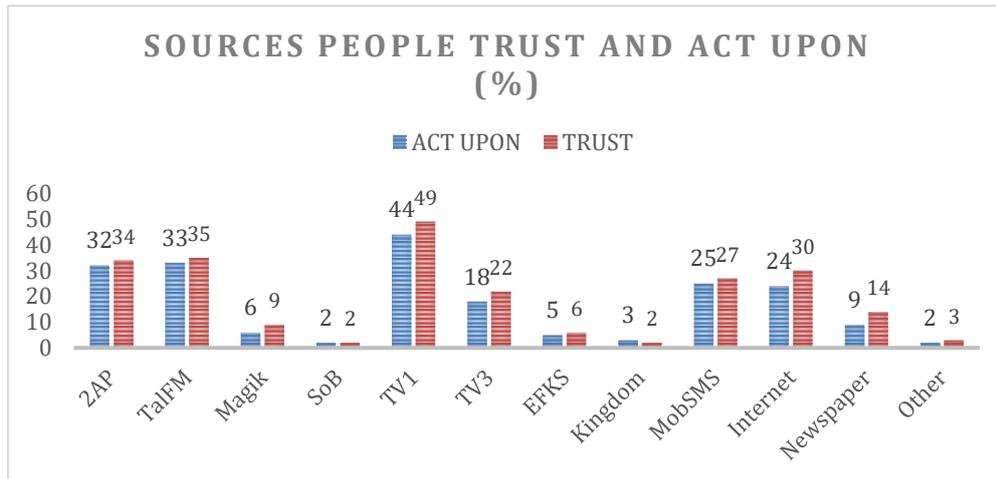

**Figure 10. What sources of information do you trust and act upon**

### 4.6.2. Whose information is most important/least important

The survey also rated whose information and what sources were the most and the least important. Results indicated that on the subject of whose information was rated the most important in an emergency, 47% indicated professional reporters followed by government at 31% and family matai (18%). It is interesting to note that information from friends was rated very low. This is confirmed by the probe on whose information is rated as the least important where there was a range of responses but 72% rated information from friends as being the least important. These ratings are very similar to those of the 2015 survey.

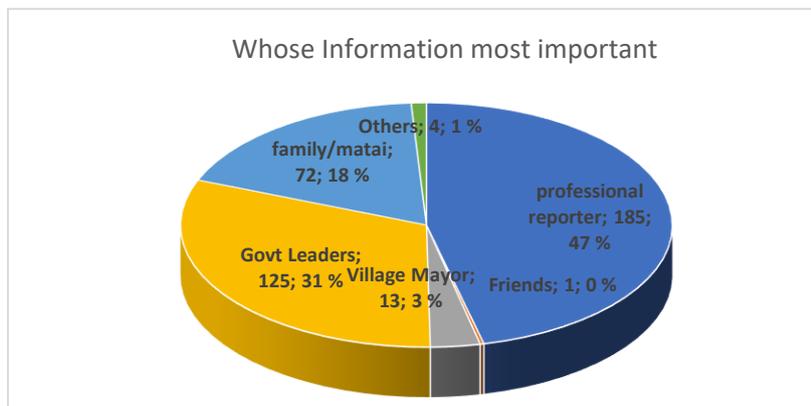

**Figure 11. Whose information is most important**

### 4.6.3. What source of Information is most important/least important

The probe on what source of information was rated as the most important, indicated that radio collectively, was rated as the most important category, particularly the government owned 2AP (39%) and Talofa FM radio station (16%) followed by TV1 television (30%). On the probe on what source was the least important, Newspapers (42%) and Internet (29%) were rated as the least important source of information in an emergency.

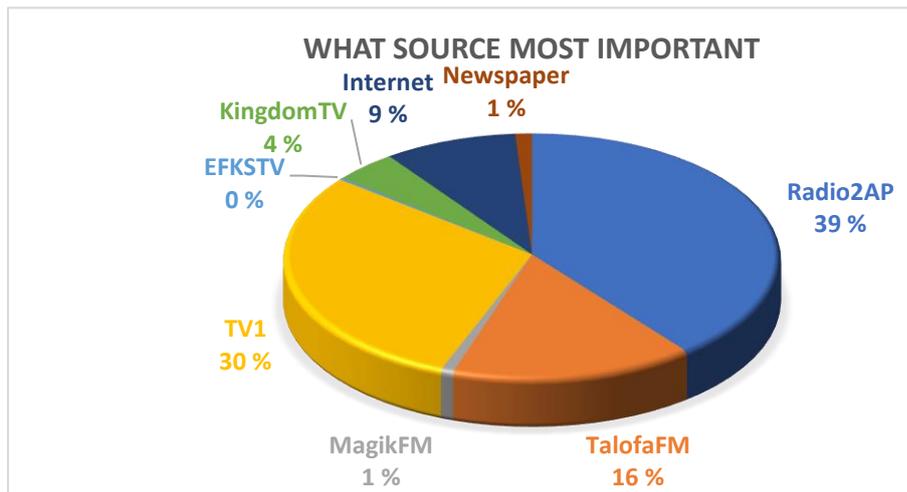

**Figure 12. What source of information is the most important**

### 4.7. Summary of Findings and Recommendations

The results summarised below provide answers to the two research questions which are the focus for this research.

The first list of findings attempts to address the first research question: **How can Samoans use their mobile and internet devices during times of crises and natural disasters?**

1. The predominance and pervasiveness of mobile phones at 94% indicate its potential for use in any disaster response and relief interventions and according to the first model can be used for lateral dissemination of information in an emergency.

2. With 82.5% of mobile phones being Smart phones – disaster preparedness, early warning and response interventions can now use not just text based but also smart apps. Mobile phones can be used by citizens to disseminate early warning, keep informed of any disaster and response updates in times of need.

3. The low usage of Twitter (11%) indicates that currently it is not the most effective media for disaster response management and lateral dissemination. However, with the huge increase in usage of IPChat applications (88%) as well as 90% Facebook usage, such social media should be considered as viable avenues for lateral dissemination for early warning and disaster response.

The rest of the findings attempts to address the second research question: **How do Samoan citizens act on different types of information and how does the source of information affect their decisions to act?**

1. People continue to place a high level of trust in both government (95%) and village leaders (84%) (matai) to provide them with the necessary emergency relief during disasters.

2. Professional reporters are regarded by the majority as the most predominantly used, the most trusted people whose information was most likely to be acted upon, and the most important source of information and news during an emergency. Professional reporters refer to official media personnel such as radio, television and newspaper reporters thus implying that radio and television are not only the most trusted sources, but also the most trusted media of information. In addition to professional reporters, people also trust and act upon information from government and family matai or chiefs. Hence this has implications on what effective sources of information need to be used for disseminating information during an emergency. Professional reporters as well as family chiefs (matai) are then reliable third party that can be used for dissemination of information in times of emergency. This supports the claim that any disaster warnings need to be issued through 'credible' sources (Ronan and Johnston 2005)
3.     People will act upon information from people sources they trust as well as media sources that they trust.
4.     The predominance of radio and TV as an important, trusted source of information which people use and most likely to act upon, indicate their important role in any emergency response interventions. As well there is also increasing usage and trust in Mobile SMS and Internet. Since these media are trusted sources they should be used extensively for dissemination of information for preparedness, early warning and disaster response.
5.     An interesting finding is the low level of importance people place on the Internet and newspapers. This may be due to people's perceptions of the lack of reliability of the Internet and newspapers, which are possibly due to these being perceived as sources of "fake news" and "delayed updates".

As mentioned earlier, the motivation to replicate the study was to evaluate whether there were any significant changes to the findings of the 2015 study. Specifically, the focus was to determine any changes in what technology devices can people use in times of disaster as well as sources and media people trust and use in emergency and disaster. A summary of the comparisons between the two studies appear below:

1.     Findings indicate that mobile phone ownership remain relatively high and similar to the 2015 figure of 91% but a massive increase in ownership of smartphones from 34% in the 2015 survey to 82.5% in the 2020 survey. Digicel was still the dominant provider but Vodafone (previously Bluesky) had now doubled its market share compared to 15% in 2015. Findings indicate a sizable increase in Internet access by mobile phone from 52% in 2015 to 89% in 2020.

2.     A huge increase (almost triple) was evident in the use of IPchat applications from 26% in 2015 to 88%, as well as 65% in 2015 to 90% in 2020 for Facebook usage. The large proportion of IPchat application users and Facebook users indicate that these types of social media can be used for dissemination of disaster warnings and response information.

3.     In terms of who to trust in times of emergency, findings indicate an increase in level of trust in both government (95%) and matai (84%) from the 2015 survey but also that people continue to place a high level of trust in government to provide them with the necessary emergency relief during disasters, as well as the village leaders (Matai). When asked as to whom they turn to for help, the most common responses were family (31%) followed by government (13%). In terms of whose information can be trusted in emergencies, findings were similar to the 2015 study in terms of trust in information from professional reporters and matai but there was also now increasing trust in government. In terms of sources people trust and act upon in an emergency, findings indicate predominance of both radio and tv as in the 2015 study, but also increasingly now the Internet and Mobile SMS This suggests that people will act upon information from sources of media that they trust.

But how do these findings contribute towards early warning and disaster response and relief management? One way to answer this is by reviewing work by Meier (2013) which indicate that in the field of disaster management, traditional centralized and external modes of early warning and response are becoming less and less effective owing to the increasing complexity of humanitarian emergencies. This has led in a shift to people centred early warning and disaster response. The 2006 UN Global Survey of (disaster) Early Warning Systems (UNISDR, 2006) defined the purpose of people-centred early warning as "to empower individuals and communities threatened by hazards to act in sufficient time and in an appropriate manner so as to reduce the possibility of personal injury, loss of life, damage to property and the environment, and loss of livelihoods". In the context of people centred early warning and disaster response, the findings from this survey have indicated which information sources are important, what information sources people trust and act upon. An example of "people centredness" is the use of mobile phones by citizens to find shelters, avoid hazardous areas, call for rescue so people can act in timely manner to avoid losses. Such information would also enable policy makers to establish a people centred early warning systems and disaster response which will empower individuals to act in sufficient time and appropriate manner in times of emergency and disaster.

Secondly, two areas from Mancur Olson's work (1965) on collective action problems which are key in a crisis situation are group size and group cohesion. ICTs such as radio, television and mobile because of their technical attributes, take care of the problems related to population size, cost, and regularity of information transmission. The findings of this survey have identified which technologies or ICTs are predominant in terms of sources people trust and will act upon. From the findings of this survey the key ICT technologies people trust and act upon are radio and television but the predominance of mobile technology in Samoa also needs to be factored in, into any early warning and disaster response systems.

In conclusion, findings from this research need to be made available to Ministry of ICT, National Emergency Operations Centre (NEOC), Telecom and Internet providers and the Disaster Management office (DMO) to inform policy making and planning. Regular updates of this information are also essential to inform optimal methods of implementing early warning and disaster response systems.

Data and findings from this study is theoretically valuable for researchers working on e-governance, while also being practically useful for external agencies such as UNDP and UNOCHA since it demonstrates the level of information the citizenry has about communication technology-supported crisis response programming, which can be an indicator of their willingness to participate and make technology-aided e-governance programs sustainable.

Such a study is also timely in the light of the current pandemic with its heavy reliance on technology and also the need to investigate how information can be transmitted vertically and laterally in times of such a disaster. It can be argued that these same types of people or community based early warning systems can be developed for pandemics and diseases such as malaria (Macherera & Chimbari, 2016).

## 5. REFERENCES AND CITATIONS